\documentclass[12pt,preprint]{aastex}

\usepackage{epsfig}

\slugcomment{}

\shorttitle{Clustering evolution of starbursts}
\shortauthors{Farrah et al.}

\begin{document}

\title{The spatial clustering of ultraluminous infrared galaxies over $1.5<z<3$}

\author{D. Farrah}
\affil{Department of Astronomy, Cornell University, Ithaca, NY 14853, USA}

\author{C. J. Lonsdale}
\affil{Infrared Processing Analysis Center, California Institute of Technology, Pasadena, CA 91125, USA}

\author{C. Borys}
\affil{Department of Astronomy and Astrophysics, University of Toronto, Toronto, Canada}

\author{F. Fang}
\affil{Infrared Processing Analysis Center, California Institute of Technology, Pasadena, CA 91125, USA}

\author{I. Waddington \& S. Oliver}
\affil{Astronomy Center, University of Sussex, Falmer, Brighton, UK}

\author{M. Rowan-Robinson \& T. Babbedge}
\affil{Astrophysics Group, Imperial College, London SW7 2BW, UK}

\author{D. Shupe}
\affil{Infrared Processing Analysis Center, California Institute of Technology, Pasadena, CA 91125, USA}

\author{M. Polletta \& H. E. Smith}
\affil{Center for Astrophysics and Space Sciences, University of California at San Diego, La Jolla, CA 92093, USA}

\author{J. Surace}
\affil{Infrared Processing Analysis Center, California Institute of Technology, Pasadena, CA 91125, USA}



\begin{abstract}
We present measurements of the spatial clustering of galaxies with stellar masses $\gtrsim10^{11}M_{\odot}$, 
infrared luminosities $\gtrsim10^{12}L_{\odot}$, and star formation rates $\gtrsim200M_{\odot}$yr$^{-1}$ in two 
redshift intervals; $1.5<z<2.0$ and $2<z<3$. Both samples cluster very strongly, with spatial correlation 
lengths of $r_{0}=14.40\pm1.99 h^{-1}$Mpc for the $2<z<3$ sample, and $r_{0}=9.40\pm2.24 h^{-1}$Mpc for the $1.5<z<2.0$ 
sample. These clustering amplitudes are consistent with both populations residing in dark matter haloes with masses 
of $\sim6\times10^{13}M_{\odot}$, making them among the most biased galaxies at these epochs. We infer, from this and previous results, that 
a minimum dark matter halo mass is an important factor for all forms of luminous, obscured activity in galaxies at 
$z>1$, both starbursts and AGN. Adopting plausible models for the growth of DM haloes with redshift, then the haloes 
hosting the  $2<z<3$ sample will likely host the richest clusters of galaxies at z=0, whereas the haloes hosting the 
$1.5<z<2.0$ sample will likely host poor to rich clusters at z=0. We conclude that ULIRGs at $z\gtrsim1$  signpost stellar 
buildup in galaxies that will reside in clusters at $z=0$, with ULIRGs at increasing redshifts signposting the buildup of 
stars in galaxies that will reside in increasingly rich clusters.
\end{abstract}


\section{Introduction} \label{intro}
Among the core principles of current theories for the formation of large-scale structure is the premise that the evolution of 
the total mass distribution can be described by the evolution of Gaussian primordial density fluctuations, and that this 
evolution is traced by galaxies in some determinable way. Overdense `halos' in the dark matter distribution are predicted to 
undergo successive mergers over time to build haloes of increasing mass, with galaxies forming from the baryonic matter in 
these haloes. This framework of `biased' hierarchical buildup \citep{col,gran,hatt} has proven to be remarkably adept in 
explaining many aspects of galaxy and large-scale structure formation. 

Comparing these predictions with observations is an active topic, and is particularly controversial concerning the evolution 
of massive galaxies (those with masses $\gtrsim10^{11}M_{\odot}$). The `naive' expectation might be that massive galaxies 
form over a long period of time, as many halo mergers are needed to build up large baryon reservoirs, and indeed some massive 
galaxies do appear to form in this way \citep{van,tam,bel}. There is however observational evidence that many massive 
galaxies may form at high redshift and on short timescales \citep{dun,ell,blak}, implying that the stars in massive local 
galaxies formed in less than a few Gyr of each other at $z>1$. Further evidence has come from blank-field sub-mm surveys 
\citep{hug,barg,eal,scot,bor,mor}, which have found a huge population of sub-mm bright sources (SMGs) at $z\gtrsim1$, with 
IR luminosities of $>10^{12}$L$_{\odot}$. SMGs are far more numerous than sources in the local Universe with comparable IR 
luminosities, the so-called Ultraluminous Infrared Galaxies (ULIRGs). These distant ULIRGs are good candidates for massive galaxies caught in the act of 
formation; they have a median redshift of $z\simeq2.5$ \citep{cha05}, high enough star formation rates to satisfy the 
observed color-scaling relationships in massive evolved systems, and a comparable comoving number density to local 
massive ellipticals. SMGs however pose a challenge for structure formation models, which invoke a diverse variety of 
solutions to explain the observed sub-mm counts \citep{vank,gran2,bau,bow}.

Progress on these issues requires observations that can be compared to models in a meaningful way, and one of the best ways 
is via measurement of clustering amplitudes. Massive haloes are predicted to cluster together on the sky, with the strength 
of clustering depending on their mass \citep{kai,bard}. If distant ULIRGs are the formation events of massive galaxies, then 
they should trace this clustering. Measuring clustering amplitudes for distant ULIRGs is however difficult; it 
requires a minimum of several hundred sources found over a significant area of sky (e.g. \citealt{vank}), but IR 
observatories available up to now cannot map large enough areas to the required depths to find this number of sources at 
$z\gtrsim1$. 

The launch of the Spitzer Space Telescope \citep{wer} offers the potential to overcome these problems, due to its ability to 
map large areas of sky in the infrared to greater depths than any previous observatory.  In this letter, we study the 
clustering of starburst-dominated ULIRGs in two redshift bins centered at $z\sim1.7$ and $z\sim2.5$. We assume $H_{0}=70$ 
km s$^{-1}$ Mpc$^{-1}$, $\Omega=1$, and $\Omega_{\Lambda}=0.7$.

\section{Analysis}\label{analysis}
To select high redshift, IR-luminous star-forming galaxies, we use the 1.6$\mu$m emission feature, which arises due to photospheric emission 
from evolved stars. When this feature is redshifted into one of the IRAC channels then that channel exhibits a peak, or 
`bump' compared to the other three channels \citep{sim,saw}. A complete discussion of the source selection and characterization 
methods is given in Lonsdale et al 2006 (in preparation), which we summarize here. 

Our sources are taken from three fields observed as part of the Spitzer Wide Area Infrared Extragalactic Survey (SWIRE, \citealt{lon}); 
the ELAIS N1 and ELAIS N2 fields, and the Lockman Hole, covering 20.9 square degrees in total. We first selected those sources 
fainter than R=22 (Vega), and brighter than 400$\mu$Jy at 24$\mu$m. Within this set, we selected two samples that displayed a 
`bump' in the 4.5$\mu$m and 5.8$\mu$m channels, i.e. where $f_{3.6}<f_{4.5}>f_{5.8}>f_{8.0}$ for one sample (the `B2' sample) and where 
$f_{3.6}<f_{4.5}<f_{5.8}>f_{8.0}$ for the other sample (the `B3' sample). This resulted in a total of 1689 B2 sources and 1223 B3 sources. 

For both samples we used the {\sc Hyper-z} code \citep{bol} to estimate redshifts, the results from 
which place most of the B2 sources within $1.5<z<2.0$, and most of the B3 sources within $2.2<z<2.8$. From the best fits we also 
derived IR luminosities and power sources, both of which are uncertain due to the photometric nature of the redshifts. 
The requirement that the sources have $f_{24}>400\mu$Jy however demands an IR luminosity of $\gtrsim10^{12}$L$_{\odot}$ for 
all the sources. In most cases the SED fits predict that the dominant power source is 
a starburst, with star formation rates of $\gtrsim200$M$_{\odot}$yr$^{-1}$ (assuming a standard Salpeter IMF). Similarly, the 
presence of the 1.6$\mu$m feature demands (accounting for uncertainties in redshift and stellar models) a 
minimum mass of evolved stars of $\sim10^{11}$M$_{\odot}$. Both the B2 and B3 sources are thus good candidates for being 
moderately massive galaxies harboring an intense, obscured starburst, making them similar in nature to both local ULIRGs, and 
high-redshift SMGs. 

We measure the angular clustering of both samples using the standard parametrization for the probability, $p$, of finding a 
source in a solid angle $\Omega_{1}$, and another object in another solid angle $\Omega_{2}$ separated by an angle $\theta$:

\begin{equation}
\label{equ:ptheta}
p(\theta) = N^2[1+\omega(\theta)]\Omega_{1}\Omega_{2},
\end{equation}

\noindent where $N$ is the mean surface density of objects and $\omega(\theta)$ is the angular two-point correlation function.  
We compute $\omega(\theta)$ using the \citet{las} prescription, and divide $\omega(\theta)$ by the "integral constraint'' 
\citep{inf} to correct for the finite size of the sample (though in both cases this correction was negligible). 
The survey area geometries are accounted for in the analysis by constructing coverage masks of the joint MIPS 24$\mu$m and 
IRAC observations, and only considering those sources that lie within these masks. 

\begin{figure*}
\begin{minipage}{150mm}
\epsfig{figure=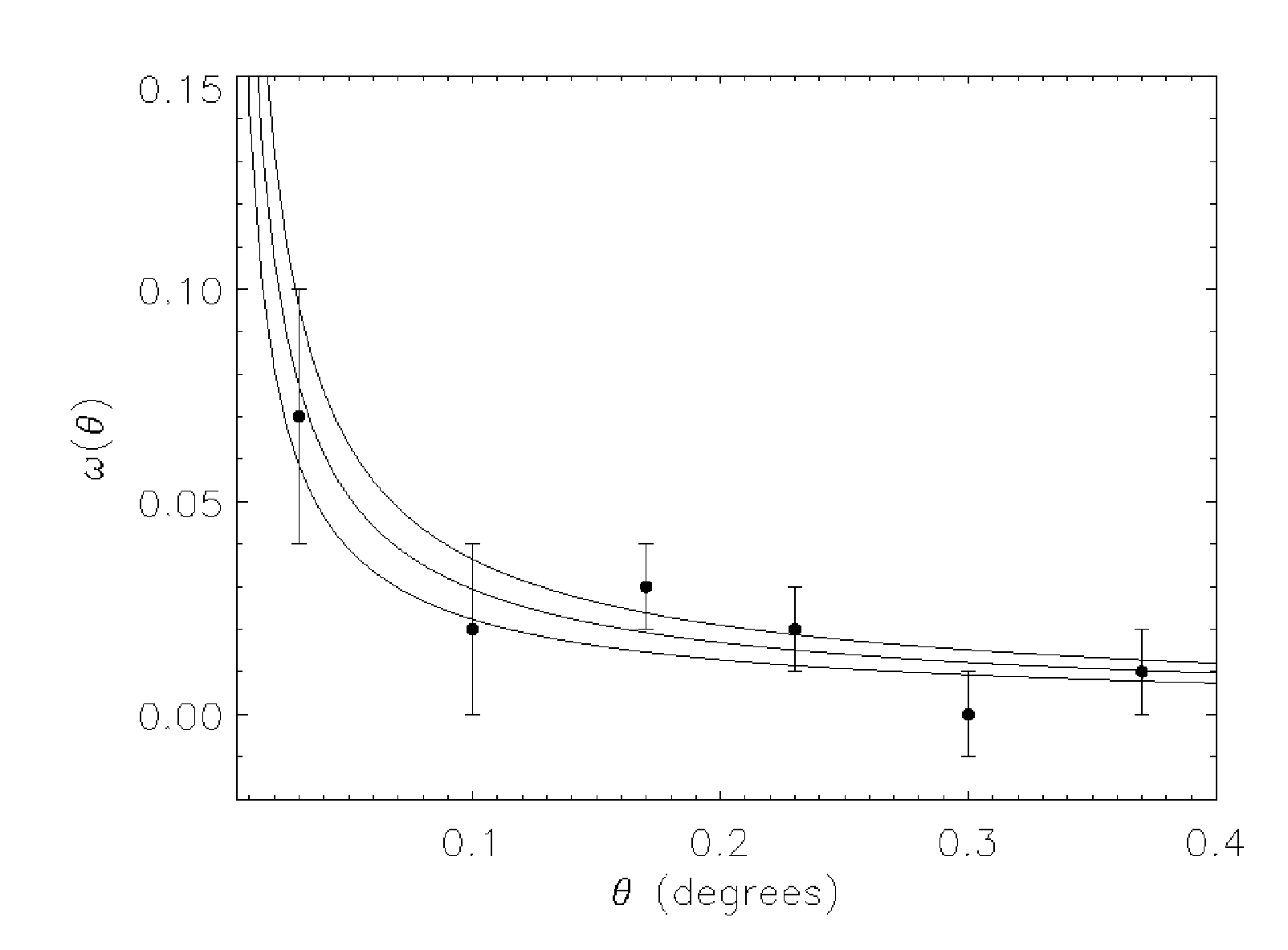,width=80mm}
\epsfig{figure=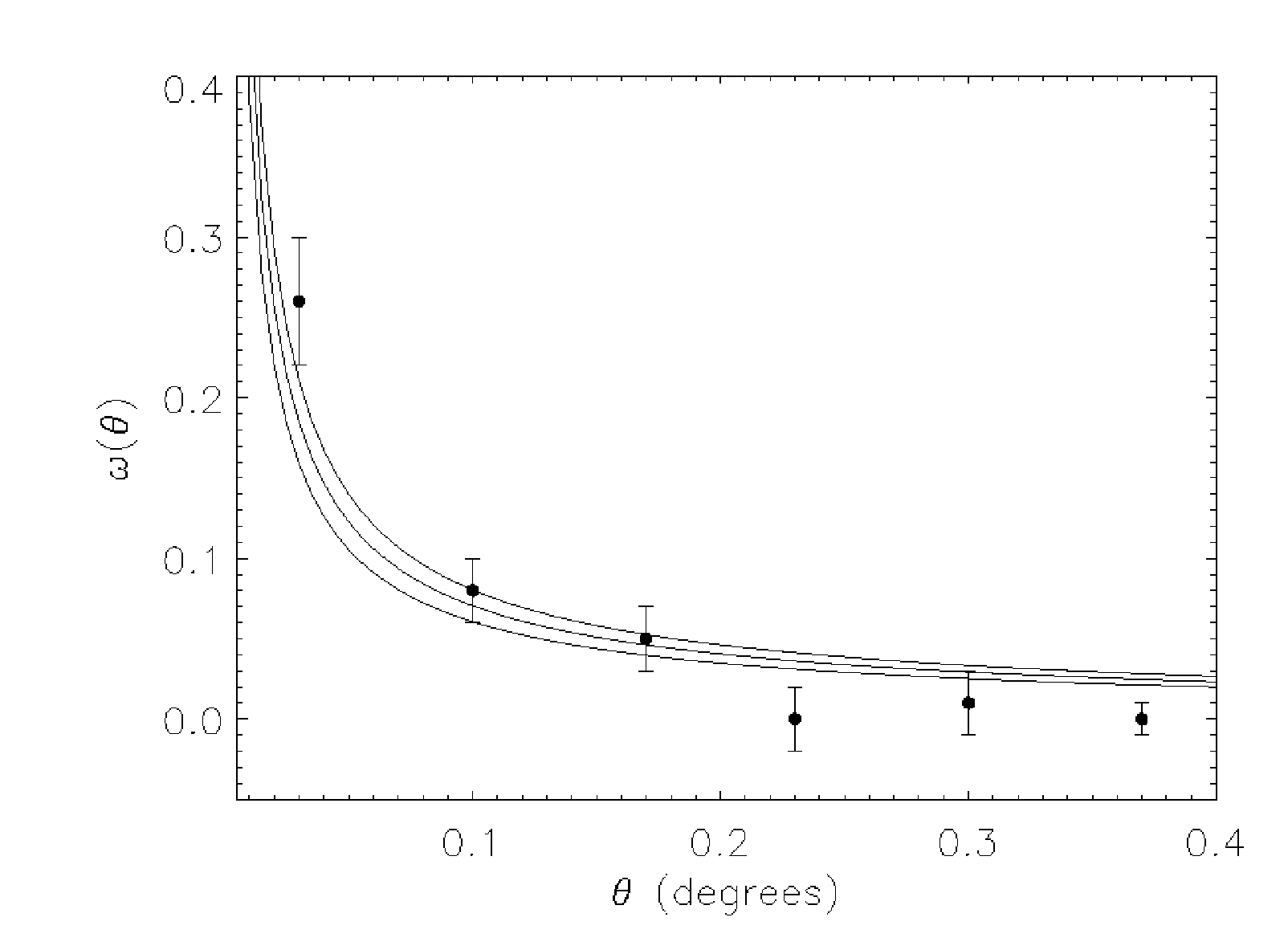,width=80mm}
\end{minipage}
\caption{The two-point correlation function and (Poisson) error bars as a function of $\theta$ for (left) the B2 sample, and 
(right) the B3 sample, in 4$\arcmin$ bins. The three lines show the best-fit power law, and the 1$\sigma$ 
limits. \label{cluster}}
\end{figure*}

We found that the levels of angular clustering seen in the three fields were consistent with each other to within 0.5$\sigma$, 
with all three fields showing positive clustering for both the B2 and B3 samples, We therefore combined the angular 
clustering measures for each sample over the three fields. In Figure \ref{cluster} we present the combined $\omega(\theta)$ 
(with Poisson errors) vs. $\theta$ plots for both samples. Both samples show a statistically significant detection of positive 
clustering on length scales of $<0.1\degr$. To quantify the strength of clustering, we fit both datasets with a single power law:

\begin{equation} \label{equ:plaw}
\omega(\theta) = A_{\omega}\theta^{1-\gamma}
\end{equation}

\noindent where $A_{\omega}$ is the clustering amplitude. A wide variety of sources are 
well fitted with $\gamma=1.8$ on small length scales \citep{bah,ove,wils}. We first check that our two datasets can be fit this way by allowing 
$\gamma$ to vary and find that $\gamma=1.8$ is within 1$\sigma$ of the best-fit values. We therefore fix $\gamma=1.8$ and fit to get 
$A_{\omega}=0.0125\pm0.0017$ for the B3s and $A_{\omega}=0.0046\pm0.0011$ for the B2s.

\section{Discussion}\label{disc}
If a sample of galaxies with a measured $\omega(\theta)$ also have a constrained redshift distribution, then we can derive the 
spatial correlation length, $r_{0}(z)$, of the sample by inverting Limbers equation \citep{lim}. For a redshift range spanning 
$a$ to $b$, this can be expressed as:

\begin{equation} \label{equ:spatc3}
\frac{r_{0}(z)}{f(z)}=\left[\frac{H_{0}^{-1}A_{\omega}cC\left[\int_{a}^{b}\frac{dN}{dz}\,dz\right]^{2}}{\int_{a}^{b}\left(\frac{dN}{dz}\right)^{2}E(z)
D_{\theta}^{1-\gamma}(z)f(z)(1+z)dz}\right]^{\frac{1}{\gamma}}
\end{equation}

\noindent where $f(z)$ parametrizes the redshift evolution of $r_{0}$, $D_{\theta}(z)$ is the angular diameter distance, $c$ is the 
speed of light, $dN/dz$ is the sample redshift distribution, and:

\begin{equation} \label{equ:spatc4}
C = \frac{\Gamma(\gamma/2)}{\Gamma(1/2)\Gamma([\gamma-1]/2)}, E(z) = [\Omega_{m}(1+z)^{3} + \Omega_{\Lambda}]^{\frac{1}{2}}
\end{equation}

To compute $r_{0}$ from Equation \ref{equ:spatc3} requires a redshift distribution, $dN/dz$, but we have no spectroscopic redshifts 
for our sample. Therefore, we use analytic forms for $dN/dz$ derived from the photometric redshift distributions \citep{lon3}. For the 
B2 sources this is a Gaussian centered at $z=1.7$ with a FWHM of $1.0$, and for the B3 sources this is a Gaussian centered at 
$z=2.5$ with a FWHM of $1.2$. The resulting correlation lengths are $r_{0}=9.4\pm2.24h^{-1}$Mpc for the B2 sources, and 
$r_{0}=14.4\pm1.99h^{-1}$Mpc for the B3 sources. As a check, we derived both redshift distributions using 
an independent code \citep{mrr}, and found that the mean redshifts were the same, but that the FWHMs were in both cases slightly 
broader. Although, in general, broader redshift distributions result in larger clustering amplitudes, in our case the effects 
are insignificant; for example if we use the \citet{mrr} redshift distribution for the B3 sources then we obtain 
$r_{0}=16.5\pm2.30h^{-1}$Mpc. To be conservative, we use the Lonsdale et al 2006 redshift distributions. 

To place these clustering results in context, we consider two classes of model. The first parametrizes the spatial correlation 
function, $\xi$, as a single power law in comoving coordinates:

\begin{equation} \label{equ:spatcl}
\xi(r_{c},z) = \left( \frac{r_{c}}{r_{0}}\right)^{-\gamma}(1+z)^{\gamma-(3+\epsilon)}
\end{equation}

\noindent where $r_{c}$ is the comoving distance, and $r_{0}$ is the comoving correlation length at $z=0$. The comoving correlation 
length at a redshift $z$, $r_{0}(z)$, can therefore be expressed as:

\begin{equation} \label{equ:spatc2}
r_{0}(z) = r_{0}f(z), f(z)=(1+z)^{\gamma-(3+\epsilon)}
\end{equation}

\noindent where the choice of $\epsilon$ determines the redshift evolution \citep{phl,ove}. Generally, models that use this 
parametrization are those where the correlation length increases or remains constant with decreasing redshift; these can be 
thought of as tracing the clustering evolution of a given halo. Several cases are usually quoted. First is `comoving clustering', 
where haloes expand with the Universe, and $\epsilon=\gamma-3$; in this case clustering remains constant with redshift. Second 
is the family of models for which $\epsilon\geq0$, for which clustering increases with time. Examples of this family 
include (a) `stable' clustering, for which $\epsilon\simeq0$ (in this case the size of the haloes is frozen in proper coordinates, 
hence comoving expansion of the background mass distribution makes the halo clustering grow stronger), 
(b) the predicted evolution of clustering of the overall dark matter distribution, where $\epsilon\simeq\gamma-1$ \citep{car2b}, 
and $r_{0}\simeq5$ at z=0 \citep{jen}, (c) `linear' clustering, where $\epsilon=1.0$. A cautionary note to this is that detailed interpretations of 
clustering evolution from these models suffer from several theoretical flaws \citep{mos,smi}, and so should be thought of as 
qualitative indicators rather than quantitative predictions. We therefore simply use the `stable' and `linear' models as 
indicators of the possible range of halo clustering amplitude with redshift. 

\begin{figure*}
\begin{minipage}{180mm}
\epsfig{figure=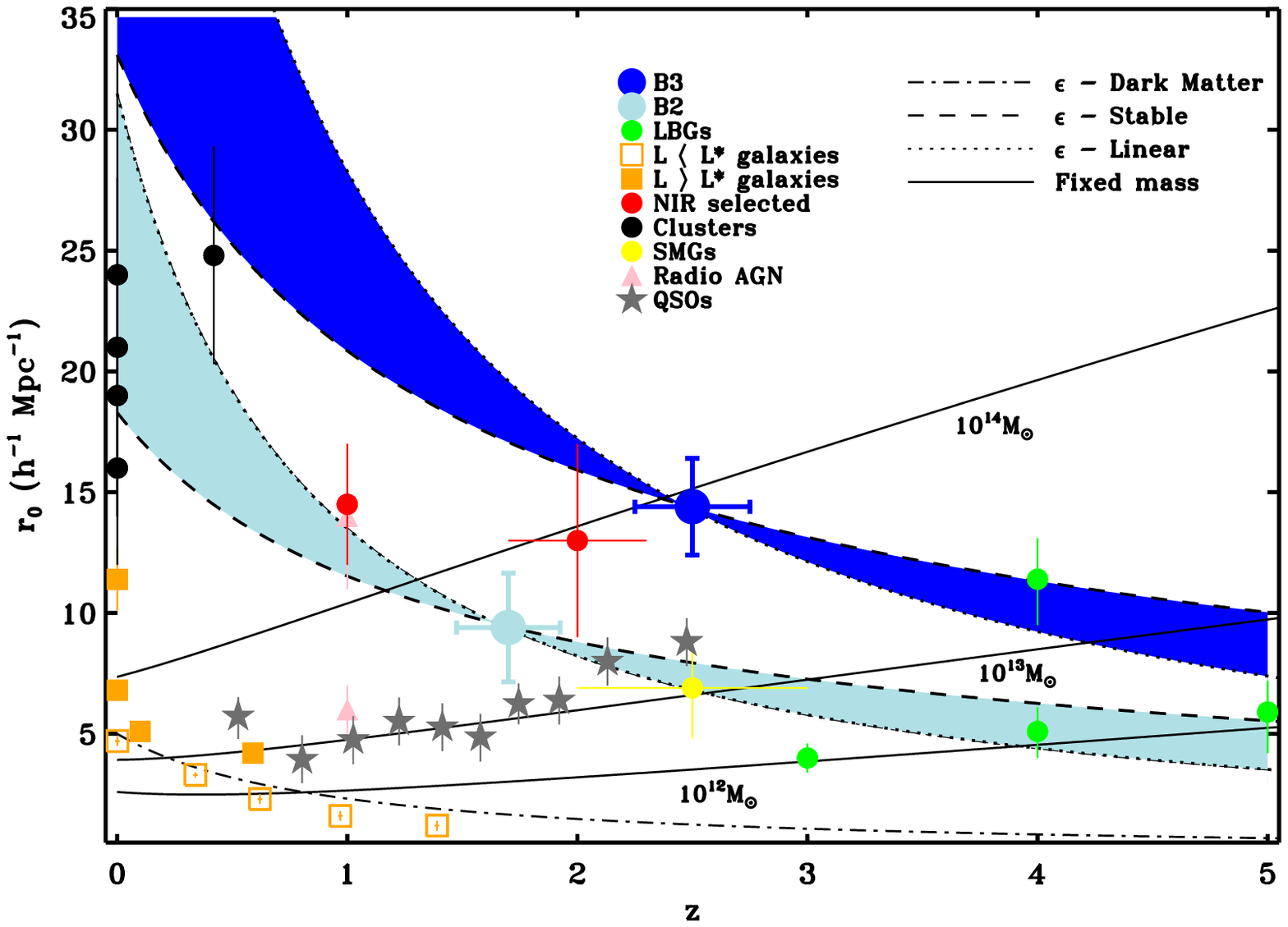,width=160mm}
\end{minipage}
\caption{Comoving correlation length, $r_{0}$, vs. redshift. Other data are taken from \citealt{mos,ove,dad,bla,ouc,ade,cro2,geo,all}. 
The `Fixed mass' lines show the predicted clustering amplitude of haloes of a given mass at any particular redshift, whereas 
the $\epsilon$ lines show the predicted clustering amplitude of an individual halo for three halo growth models, described 
in the text. The `Stable' and `Linear' lines give a qualitative indicator of the range of how DM haloes may grow with redshift, 
and we have normalized `Stable' and `Linear' lines to the clustering amplitudes of the B2s and the B3's. The shaded regions therefore 
indicate what these haloes may host at lower and higher redshifts. - the haloes hosting B3s may contain an optically bright LBG at 
$z\simeq4$ (upper green point), and grow to host very rich galaxy clusters at z=0, whereas the haloes hosting B2 sources may 
contain optically fainter LBGs at $4<z<5$, SMGs at $z\sim2.5$, radio-bright AGN (upper pink triangle) and (old) EROs at $z\simeq1$, and poor 
to rich clusters at z=0.  
\label{r0plot}}
\end{figure*}

The second class of model comprises those in which comoving correlation lengths increase with {\it increasing} redshift. These 
models introduce `bias', $b(z)$, between the galaxies and the underlying dark matter. An example of such models are the `fixed 
mass' models \citep{mata,mos}, which assume that galaxies selected in a certain way predominantly occupy haloes of the same mass. 
These models therefore predict the clustering strength of haloes of a specified mass at any given redshift. In Figure \ref{r0plot} we plot the $\epsilon$ model 
for dark matter, `stable'and `linear' epsilon models normalized to the B2 and B3 clustering strengths, the fixed halo mass models 
for halo masses of $10^{12}$M$_{\odot}$, $10^{13}$M$_{\odot}$, and $10^{14}$M$_{\odot}$, the $r_{0}$ values for the B2 and B3 
galaxies, and the spatial correlation lengths of other galaxy populations taken from the literature. 

Considering these uncertainties, care must be taken when comparing the models to observed galaxy correlation lengths. With this in 
mind, we use Figure \ref{r0plot} to explore the 
relationships between our samples, the underlying dark matter, and other galaxies. Considering first other measured or predicted 
correlation lengths at the same redshift, then both samples are very strongly clustered, with correlation lengths much 
higher than that predicted for the overall DM distribution at their respective epochs. Both B2s and B3s cluster significantly more strongly than 
optical QSOs at their respective epochs, and B3s cluster more strongly than sub-mm selected galaxies. Based on the \citet{mata} models, then we derive 
{\it approximate} 1$\sigma$ halo mass ranges of $10^{13.7}<$M$_{\odot}<10^{14.1}$ for the B3s, and $10^{13.5}<$M$_{\odot}<10^{13.9}$ 
for the B2s. 

The most interesting comparison is however between the two samples themselves. The clustering evolution of QSOs with redshift 
\citep{cro2} may mean that there is a `minimum' host halo mass for QSO activity, below which no QSO is seen, of $\sim5\times10^{12}$M$_{\odot}$. 
The correlation lengths for the B2 and B3 samples are consistent with the same conclusion but for a higher halo mass, tracing 
the clustering line for a $\sim6\times10^{13}M_{\odot}$ DM halo. Interestingly, from Figure \ref{r0plot} we would draw a similar 
conclusions for optically faint LBGs, albeit for a halo mass of $\sim10^{12}M_{\odot}$ (see also \citet{ade}). Taken together, these results imply 
that a minimum halo mass is an important threshold factor for {\it all} forms of very luminous activity in galaxies, both starbursts 
and AGN. It is also interesting to speculate on what the host haloes of B2 and B3 sources contain at lower and higher redshifts. We 
might expect that a halo hosting a B3 source could contain an optically bright LBG at $z\sim4$ when its mass is $\sim10^{13.2}$M$_{\odot}$, 
followed by a B3 at $z\sim2.5$ once the halo has reached $\sim10^{14}$M$_{\odot}$, possibly accompanied by other (near-IR selected) 
star forming systems \citep{dad,dad2}, before evolving to host an extremely rich galaxy cluster at low redshifts, with a halo mass 
substantially exceeding $10^{15}$M$_{\odot}$.  The occupants of a halo hosting a B2 galaxy would however probably be different. 
We would expect that such a halo could contain an SMG at $z\sim2.5$, and optically fainter LBGs at $4<z<5$ (though probably not 
LBGs at $z\sim3$). At lower redshifts such a halo might host a radio-bright AGN and or ERO at $z\sim1$, and a (poor to rich) 
cluster at $z=0$. These evolutionary paths for the B2s and B3s are consistent 
with semi-analytic models and numerical simulations \citep{gov,nag} which predict that LBGs at high redshift evolve into local 
clusters. We conclude that ULIRGs at $z\gtrsim1.5$ as a class likely signpost stellar buildup in galaxies in clusters at $z=0$, with 
higher redshift ULIRGs signposting stellar buildup in galaxies that will reside in more massive clusters at lower redshifts. These 
predictions are however sensitive to the assumed form of DM halo growth, and should be regarded with caution.

\acknowledgments
We thank Emanuele Daddi for saving our lives, Andy Connolly for useful discussion, and the referee for a very helpful report. SJO acknowledges 
a Leverhulme Fellowship. Support for 
this work, part of the Spitzer Space Telescope Legacy Science Program, was provided by NASA through an award issued by JPL 
under NASA contract 1407. This research was carried out, in part, by JPL, California Institute of Technology, and was sponsored by NASA.

\end{document}